\documentclass[draftcls,12pt,onecolumn,oneside]{IEEEtran}
\usepackage{mathrsfs}
\usepackage{algorithm}
\usepackage{algorithmic}
\usepackage{url,times,amsmath,color,amssymb,graphicx,epsfig,psfig,cite,geometry,psfrag,subfigure,dsfont,booktabs, multirow}
\begin{document}
\title{Cooperative Interference Mitigation and Handover Management for Heterogeneous Cloud Small Cell Networks}
\author{Haijun Zhang,~\IEEEmembership{Member,~IEEE}, Chunxiao Jiang,~\IEEEmembership{Member,~IEEE}, Julian Cheng,~\IEEEmembership{Senior Member,~IEEE},   and Victor C.M. Leung,~\IEEEmembership{Fellow,~IEEE}
\thanks{Haijun Zhang and  Victor C.M. Leung are with the Department of Electrical and Computer Engineering, The University of British Columbia, Vancouver, BC V6T 1Z4 Canada (e-mail: haijunzhang@ece.ubc.ca, vleung@ece.ubc.ca).

Chunxiao Jiang is with the Department of Electronic Engineering, Tsinghua University, Beijing
100084, P. R. China (e-mail: jchx@tsinghua.edu.cn, corresponding author).

Julian Cheng is with the School of Engineering, The University of British
Columbia, Kelowna, BC, V1V 1V7 Canada (e-mail: julian.cheng@ubc.ca).

}}
\maketitle

\begin{abstract}
Heterogeneous small cell network has attracted much attention to satisfy users' explosive data traffic requirements. Heterogeneous cloud small cell network (HCSNet), which combines cloud computing and heterogeneous small cell network, will likely play an important role in 5G mobile communication networks.  However, with massive deployment of small cells, co-channel interference and handover management are two important problems in HCSNet, especially for cell edge users. In this article, we examine the problems of cooperative interference mitigation and handover management in HCSNet. A network architecture is described to combine cloud radio access network with small cells. An effective coordinated multi-point (CoMP) clustering scheme using affinity propagation is adopted to mitigate cell edge users' interference. A low complexity handover management scheme is presented, and its signaling procedure  is analyzed in HCSNet. Numerical results show that the proposed network architecture, CoMP clustering scheme and handover management scheme can significantly increase the capacity of HCSNet while maintaining users' quality of service.

\end{abstract}

\section{Introduction}
Heterogeneous small cell network  has attracted much attention due to the explosive demand of users' data requirement. In heterogeneous small cell network, low power small cells (such as picocell, relay and femtocell) together with macrocells, can improve the coverage and capacity of cell-edge users and hotspot by exploiting the spatial reuse of spectrum \cite{TonyQuekCambridge2013}. Small cells can also offload the explosive growth of wireless data traffic from macrocells. For example, in an indoor environment WiFi and femtocells can offload most of the data traffic from macrocells \cite{MugenPengJSAC2015}. For mobile operators, small cells such as femtocells can reduce the capital expenditure (CAPEX) and operating expenditure
(OPEX) because of the self-installing and self-operating features of femto basestations.

Cloud radio access network (C-RAN) is one promising wireless network architecture in 5G networks, and it was first proposed by China Mobile Research Institute\cite{MugenPengIEEENet2015}. In C-RAN, baseband processing is centralized in baseband unit (BBU) pool, while radio frequency (RF) processing is distributed in remote radio head (RRH). The C-RAN network architecture can reduce both CAPEX and OPEX for mobile operators, because fewer BBUs are potentially required in the C-RAN architecture, and the consumed power is lowered \cite{CRANsurvey2014}.

The combination of heterogeneous small cell network and C-RAN, which is called heterogeneous cloud small cell network (HCSNet), benefits from employing both C-RAN and small cell networks \cite{HCRANmugenIWC2014}. First, C-RAN reduces the power and energy cost in HCSNet by lowering the number of BBUs in densely deployed heterogeneous small cell network. Second, BBUs can be added and upgraded without much effort in the BBU pool, and network maintenance and operation can also be performed easily. Third, many radio resource management functions can be facilitated in the BBU pool with little delay. In HCSNet, cloud computing enabled signal processing can be fully utilized to mitigate interference and to improve spectrum efficiency  in 5G networks.

In the literature, HCSNet has been studied extensively. In \cite{HCRANmugenIWC2014}, state-of-the-art research results and challenges were surveyed for heterogeneous cloud radio access networks, and promising key techniques were investigated to improve both spectral and energy efficiency. To mitigate the interference for cell-edge users, coordinated multi-point (CoMP) transmission/reception is also investigated in a C-RAN environment. Though there are many interference mitigation schemes,  we will only focus on the CoMP based interference mitigation in this article due to the space limitation. C-RAN architecture is an effective network architecture to implement CoMP. Energy efficient resource optimization was studied in \cite{CRANRAmugenTVT2014} for C-RAN enabled heterogeneous cellular network. In \cite{CRANCoMP2013GC}, the authors investigated the joint  transmission   CoMP  performance  in  a  C-RAN implementation of LTE-Advanced heterogeneous network with large CoMP cluster sizes. CoMP was also investigated in \cite{HCRANmugenIWC2014} to mitigate the interference in heterogeneous cloud radio access networks. Although HCSNet has advantage in data processing, efficient clustering scheme with low complexity is still required. Therefore, clustering scheme is one of the most important factors to affect the performance of CoMP. To the best of our knowledge, the clustering of CoMP in cooperative interference mitigation of HCSNet has not been well investigated.

Another important challenge for  HCSNet is seamless mobility  handover of users. Since mobility management is a wide topic, we only focus on handover management in this article. Most of the traditional handover decision schemes were based on reference signal receiving power and reference signal receiving quality. The handover procedures and signaling flows in HCSNet will be different from the traditional small cell networks \cite{CRANsurvey2014}. In \cite{MobilityCRANIEEENetw2014}, the authors analyzed the mobility handover control in cloud computing enabled wireless networks and indicated that mobility is an inherent feature of today's wireless network. The authors in \cite{HCRANmugenIWC2014} also discussed handover management in heterogeneous cloud radio access networks; however, a quantitative analysis was missing. Generally, handover management for HCSNet has received little attention in the current literature.

Different from the existing studies in HCSNet, we focus on cooperative interference mitigation based on CoMP and handover management in HCSNet. The cluster scheme of CoMP is studied in cooperative interference mitigation of HCSNet. We first present network architecture of a cloud computing based HCSNet in Section II. Then we describe in Section III an effective CoMP clustering scheme in HCSNet using affinity propagation to mitigate cell-edge users' interference. Handover management scheme, including handover procedures, is presented in Section IV. Finally, Section V concludes this article.

\section{HCSNet Architecture}

Inspired by \cite{MugenPengIEEENet2015}\cite{HCRANmugenIWC2014}, we describe a network architecture of HCSNet as shown in Fig. 1, which is comprised of both macrocells and small cells, where CoMP is deployed. As shown in Fig. 1, macro basestations  and small basestations are reduced to macro RRH (MRRH) and small RRH (SRRH) in HCSNet, respectively. The resource management and control capabilities of BBUs of macrocell and small cell are co-located and processed in the BBU pool.  A BBU pool  consists of general-purpose processors that perform baseband processing. Different BBU pools are connected by X2 interfaces. RRHs are located  in different sites to provide wireless signal coverage for the user equipment (UE).
Millimeter-wave radio is used for fronthaul links (between BBU pools to RRHs) in HCSNet.

RRHs can be deployed on each floor of a building or office to provide enhanced coverage and capacity. RRHs can also be deployed in a hotspot scenario, e.g., stadium. Though the interference can be high, cooperative interference management will be efficiently implemented in HCSNet architecture. The BBU Pool usually  supports 100 MRRHs for a medium sized urban network (coverage 5x5 km), 1000 MRRHs for 15x15 km \cite{CRANsurvey2014}. The number of the SRRHs will be much more than MRRHs for the same size mentioned above and it depends on the specific scenarios.

In the scenario of HCSNet with high mobility users, handovers may frequently happen between small cells because of small cell size. Handover management is therefore essential and it will be processed in the BBU pools.


\section{Cooperative Interference Mitigation using CoMP in HCSNet}

In HCSNet, deployment of small cells will result in co-channel interference.  Thus, interference management is essential in HCSNet. Since signal processing is centralized in the BBU pool of HCSNet, delay is reduced in processing and transmitting. Therefore, interference mitigation such as enhanced inter-cell interference coordination (eICIC) and CoMP are greatly facilitated.

\subsection{Cloud CoMP Architecture for Interference Mitigation}

CoMP technique has been proposed as a key technique to mitigate cell-edge users' inter-cell co-channel interference. Together with HCSNet, CoMP can improve system coverage and cell-edge user spectral efficiency in future 5G networks. There are four different CoMP strategies: joint  transmission, dynamic point selection, dynamic point blanking, and coordinated scheduling/beamforming \cite{HaijunTVT2014}. In this article, we will only focus on joint  transmission in which several RRHs form a coordinated RRH cluster  and jointly serve the UEs (see Fig. 2). The RRHs in a coordinated RRH cluster are connected to the cloud via fronthaul links on which control signaling and user data are exchanged. There are two kinds of RRH clusters are involved: measurement RRH cluster and coordinated RRH cluster. Measurement RRH cluster denotes the set of RRHs which share measurement information such as power levels and channel state information (CSI), while coordinated RRH cluster denotes a set of RRHs that jointly receive and process data from the CoMP user. Note that, we don't differentiate MRRH and SRRH in Section III.

To reduce the fronthaul overhead, several clustering schemes for coordinated RRH cluster have been proposed in the literature. The existing measurement RRH cluster and coordinated RRH cluster clustering strategies can be classified into three categories: static
clustering, semi-dynamic clustering and full-dynamic clustering \cite{HaijunTVT2014}.  The static scheme, which doesn't require dynamic information exchange, is simple to operate but it can only provide limited throughput
gain \cite{SimCoMP2010IWC}. Measurement RRH cluster is identical to coordinated RRH cluster in static clustering strategy and is fixed by the network. While in full-dynamic and semi-dynamic schemes, the coordinated RRH cluster is a subset of measurement RRH cluster. Though full-dynamic clustering schemes can mitigate co-channel interference dynamically, the large signaling flow  in dense HCSNet cannot be ignored. Because the size of measurement RRH cluster in full-dynamic scheme is usually very large. Therefore, the information collection of signal strength and channel state information (CSI) of measurement RRH cluster is heavy. That is, full-dynamic schemes can achieve good performance at the expense of exhaustive information interchange/collection. A semi-dynamic clustering scheme, which offers a balance between performance and complexity, chooses measurement RRH cluster without using dynamic channel information but determines coordinated RRH cluster using dynamic channel information.

In \cite{CRANCoMP2013GC}, the authors  analyzed  the  system-level performance  of  joint  transmission  CoMP  in  a  C-RAN  implementation  of LTE-Advanced  heterogeneous networks. However, the method in \cite{CRANCoMP2013GC} requires full CSI. In HCSNet CoMP, small cells are usually densely deployed, where complete CSI is hardly available. Affinity propagation is a  fast
converging iterative algorithm \cite{APScience2007} that is suitable for CoMP RRH selection of coordinated RRH cluster which only provides limited prior information \cite{CoMP2012AP}.

\subsection{AP Clustering Based Cloud CoMP Scheme in HCSNet}

Considering the tradeoff between performance and complexity, we present in this subsection a semi-dynamic clustering framework consisting of offline and online phases to
maximize the spectrum efficiency and throughput with low fronthaul traffic in CoMP enabled HCSNet. Measurement RRH cluster for CoMP is determined based on geographical locations and the reference signal received power during the offline phase; and for the online phase we describe a clustering algorithm to choose coordinated RRH cluster from measurement RRH cluster \cite{HaijunTVT2014}. The affinity propagation  principle is employed during the online phase and we name the presented clustering scheme as affinity
propagation based clustering (APBC). Compared to existing static and full-dynamic clustering schemes, the presented semi-dynamic scheme is effective and only requires limited CSI between the local and the neighboring cells.

In Fig. 3(a), we decompose the semi-dynamic clustering
scheme into two phases. The offline phase identifies the measurement RRH cluster based on geographical location and reference signal received power, while the online
phase chooses the coordinated RRH cluster from measurement RRH cluster. The detailed clustering procedure is  described as follows.

In practical systems, only limited number of RRHs can cooperate in order to make communication overhead affordable.
The network launches offline phase and forms a measurement RRH cluster for the user based on geographical locations of RRHs and reference signal received power. For the online phase, the UE's received signal-to-interference-plus-noise ratio (SINR) of the sounding reference signal sent by  RRHs can be fed back to RRHs periodically.
A lower SINR of sounding reference signal indicates that the user suffers greater interference and requires CoMP service.

APBC takes an input key variable called ``similarity" and  maintains two information variables called ``responsibility" and ``availability". The output of AP is ``exemplar" and its associated nodes, forming a coordinated RRH cluster. The exemplar for RRH $i$ represents the master RRH of the cluster including RRH $i$.
The exemplar represents the master RRH of RRH $i$. The \emph{similarity} $s(i,k)$ indicates how well RRH $k$ is suitable to be the exemplar for RRH $i$. In particular, $s(k,k)$ is referred to ``preference", and a RRH with larger \emph{preference} values is more likely to be chosen as the exemplar. The \emph{similarity} matrix is a unique input of the APBC algorithm and has a direct impact on the performance. In this work, we define the off-diagonal elements of \emph{similarity} matrix based on pair CoMP SINR gain ($pcg$), which is defined as the ratio of estimated SINRs of the pairwise CoMP and non-CoMP RRH, i.e., $pcg = SIN{R_{CoMP}}(k,i)/SIN{R_{non - CoMP}}(k,i)$.

The \emph{responsibility} $r(i,k)$ is sent from RRH $i$ to candidate exemplar RRH $k$ as shown in Fig. 3(b) \cite{APScience2007}.
and it reflects how well RRH $k$ serves as the exemplar RRH for RRH $i$ by taking into consideration of other potential exemplars for RRH $i$.
\emph{Availability} is sent from candidate exemplar RRH  to RRH $i$ as shown in Fig. 3(c) \cite{APScience2007}.
\emph{Availability} reflects the accumulated evidence that how appropriate  RRH $i$ chooses RRH $k$ as its exemplar.

\subsection{Evaluation of AP Clustering Based Cloud CoMP Scheme in HCSNet}

The spectrum efficiency cumulative distribution function (CDF) curves of edge UEs  are plotted in Fig. 4(a). It can be observed that the four studied CoMP schemes achieve higher spectrum efficiency than the non-CoMP scheme; therefore, CoMP can efficiently combat the inter-cell interference in HCSNet. Moreover, APBC scheme provides better spectrum efficiency for edge users than the other schemes such as the static CoMP, signal-interference matrix based CoMP clustering scheme (sim-CoMP)  \cite{SimCoMP2010IWC} and Wesemann's scheme \cite{CoMP2012AP}. This is because the effective information interchange mechanism and input of $pcg$ are designed in our presented semi-dynamic clustering APBC algorithm.

Figure 4(b) compares the run time of static-CoMP, sim-CoMP, Wesemann's scheme and the APBC scheme. The run time refers to the executing time of one algorithm in this article.
We observe that when the number of RRHs increases, the run time of all four considered schemes increases as expected. However, the run time of APBC CoMP scheme is the lowest when the number of small cells is greater than 19. This is because APBC's clustering scheme is semi-dynamic as shown in Fig. 3(a), and the presented algorithm can be implemented in each cluster using only local information and limited CSI between neighboring RRHs. Thus, APBC CoMP scheme can achieve higher throughput with lower computational complexity.

The complexity of the affinity propagation algorithm in terms of run time depends on the number of iterations. If there are $n$ RRHs and there are $n^2$ values in the responsibility matrix $R$, then the updating responsibility $r(i,j)$ requires $O(n-1)$ and the updating availability $a(i,j)$ takes $O(n-2)$  for each value. As a result, each iteration requires $O(2*n^3-3*n^2)=O(n^3)$. Moreover, the convergence of affinity propagation algorithm has been proven in \cite{CoMP2012AP}, which can guarantee the practicality of the presented algorithm.

\section{Handover Management in HCSNet}

Handover management is one of the key techniques to satisfy users' quality of service (QoS)  requirement in mobile communications. However, handover management in HCSNet has not received enough attention in the existing literature. In \cite{HCRANmugenIWC2014}, the authors conducted a survey for HCSNet and discussed how high-mobility UEs should be served by macrocells with reliable connections and the low mobility UEs should be served by SRRHs. In densely deployed HCSNet, handovers occur frequently, causing heavy burden to fronthaul and core networks. Besides, mobility handover related radio link failure (RLF) and unnecessary handover (e.g., ping-pong handover) may happen because of the small size of SRRH's coverage and severe co-channel interference. In C-RAN enabled HCSNet, interrupt time and delay of handover can be reduced because handover can be accomplished within the BBU pool.

\subsection{Handover Procedures in HCSNet}
Because handover management in  HCSNet architecture is different from the traditional E-UTRAN network architecture and the existing heterogeneous network architecture, the HCSNet related handover procedure should be modified in both SRRH-SRRH handover and MRRH-SRRH handover. Owing to different sizes of macrocells and small cells, MRRH-SRRH handover is more challenging than SRRH-SRRH handover. Due to the introduction of C-RAN in the considered architecture, many radio resource management functions are moved to BBU pool. Different from handover procedure between macrocells in LTE system, many handover related functions, such as handover decision and admission control in SRRH and MRRH, are moved to the BBU pool. These SRRHs/MRRHs do not support mobility management functions.  The mobility management functions of both SRRHs and MRRHs are supported by the BBU pools as shown in Fig. 1. A BBU pool contains layer 3. To the authors' best knowledge, no existing work has proposed handover procedure between the BBU pools in HCSNet.  Therefore, we introduce in Fig. 5 an inter-BBU-pool MRRH-SRRH handover call flow based on the HCSNet architecture. Handover call flow between SRRHs follows that of MRRH-SRRH handover. The handover signaling flow of intra-BBU-pool is simpler than that of inter-BBU-pool. Due to the space limitation, we only focus on the inter-BBU-pool scenario in this article.

\subsection{Inappropriate Handover Detection Method in HCSNet}
Due to the dense deployment of small cells, the number of handover related RLFs and unnecessary handovers needs to be reduced. These undesirable handovers are described as follows \cite{EurasipMobility2013}: 1) Ping-pong handover: A handover back to the serving cell from the target RRH shortly after a successful handover to the target RRH;  2) Continue handover: A handover to another RRH (neither the original serving RRH nor the target RRH) shortly after a successful handover to the target RRH;  3) Late handover: A RLF occurs under a serving RRH before handover or during the handover procedure, and then the UE reconnects to the target RRH (different from the serving RRH);  4) Early handover: A RLF occurs shortly after a successful handover to the target RRH, and
then the UE reconnects to the serving RRH;  5) Wrong handover: A RLF occurs shortly after a successful handover to the target RRH, and then the UE reconnects to another RRH (neither the serving RRH nor the target RRH).
In order for the cloud to detect one of those scenarios, the following procedure is applied. RRH starts the timer for each UE at the moment of receiving the handover completion from each UE. During the connecting time period, if the RRH receives the RLF report from other RRHs, the RRH stops the timer. Based on the performance metric definitions, according to UE's status after the RLF, RRH categorizes RLF as a call drop, too late handover, too early handover or wrong handover.

\subsection{A Low Complexity Handover Optimization Scheme in HCSNet}
The high speed macrocell UEs usually do not need to handover to the small cell while the low speed UEs may wish to
handover to the small cell. The traditional handover scheme lets the high speed macrocell users handover to the small cell,
which may introduce two times unnecessary handovers for the user. We present a simple and effective low complexity
handover scheme to optimize the system performance. The main idea of the low complexity scheme is described as follows: Suppose a handover is about to occur. A high speed user doesn't handover to SRRH. For medium speed users, users with real-time service handover to small cell and users with non-real-time service do not handover to small cell. Low speed users handover to small cell \cite{HaijunZhangVTC2011}. Note that, the UE's speed can be estimated based on Doppler spread frequency or autocorrelation function \cite{ICCCSpeed2014}.

In order to verify the performance of the optimized schemes presented in this subsection, we compare the optimized scheme
 with traditional handover scheme in terms of system signalling overhead in different scenarios.

Figure 6(a) shows the signalling overhead (which is unitless and to be proportional to the delay required to send or process a signalling message) versus the mean of the session holding time in HCSNet with the proportion of high mobility state users $\alpha$ being set to 0.1. As seen in Fig. 6(a), the total signalling overhead increases as the mean of session holding time increases. This is because that the larger the session holding time is, the bigger the probability of cell-crossing is, implying a high probability of handover.

Figure 6(b) shows results of the signalling overhead versus the proportion of the mobility users. As Fig. 6(b) shows, the number of handovers and signalling overhead in traditional scheme increases with higher $\alpha$, while the signalling overhead decreases in the optimized handover scheme. In the scheme optimized in this section, we do not allow the high speed users handover from MRRH to SRRH while low speed users are allowed. Therefore, when $\alpha$ increases close to 1, the signalling cost in the optimized algorithm decreases to zero. From Fig. 6, it can be seen that as the average session holding time increases, the signalling cost in both traditional algorithm and optimized algorithm increases, since more handovers are expected with an increase of the session holding time. In a traditional handover algorithm for SRRHs, the high speed users and low speed users are treated as the same. Two unnecessary handovers happen as the UE handover happens as the UE moves from MRRH to SRRH.
As the optimized scheme doesn't allow high speed users' two unnecessary handovers, the total cost of the handover is reduced. As a result, there is a big decrease in the optimized scheme's signalling overhead.

\section{Conclusion}
In this article, we examined cooperative interference mitigation and handover management in HCSNet, where cloud radio access network is combined with small cells.  An effective CoMP clustering scheme using affinity propagation was presented to mitigate cell edge users' interference. A handover management scheme was presented, and  handover signaling procedures were analyzed for HCSNet. Numerical results demonstrated that the proposed network architecture, CoMP cluster scheme and handover management scheme can significantly increase the capacity of HCSNet while maintaining users' quality of service. We will consider the joint interference mitigation and handover management, as well as joint time delay and clustering in the future works.
Moreover, self-organized HCSNet can also be considered as a future direction, where interference mitigation and handover management are controlled in a self-organizing way. Interference mitigation can be benefited from self-optimizing power control, and handover management can be enhanced by automatic neighbor relation and physical cell ID self-configuration in self-organized HCSNet.

\section*{Acknowledgment}
This work was supported by the National Natural Science Foundation of China (61471025, 61371079).

\begin{IEEEbiography}{Haijun Zhang} (M'13) received his Ph.D. degree from Beijing University of Posts Telecommunications (BUPT). He hold a Postdoctoral Research Fellow position in Department of Electrical and Computer Engineering, the University of British Columbia (UBC). He was an Associate Professor in College of Information Science and Technology, Beijing University of Chemical Technology. From 2011 to 2012, he visited Centre for Telecommunications Research, King's College London, London, UK, as a joint PhD student and Research Associate.  He has published more than 50 papers and has authored 2 books. He serves as the editors of Journal of Network and Computer Applications, Wireless Networks (Springer), and KSII Transactions on Internet and Information Systems. He served as Symposium Chair of GAMENETS'2014 and Track Chair of ScalCom'2015. He also serves or served as TPC members of many IEEE conferences, such as Globecom and ICC. His current research interests include 5G, Resource Allocation, Heterogeneous Small Cell Networks and Ultra-Dense Networks.
\end{IEEEbiography}

\begin{IEEEbiography}{Chunxiao Jiang}(S'09-M'13)
received his B.S. degree in information engineering from Beijing University of Aeronautics and Astronautics (Beihang University) in 2008 and the Ph.D. degree from Tsinghua University (THU), Beijing in 2013, both with the highest honors. During 2011¨C2013, he visited the Signals and Information Group at the Department of Electrical \& Computer Engineering of the University of Maryland with Prof. K. J. Ray Liu. Dr. Jiang is currently a post-doctor in EE department of THU with Prof. Yong Ren. His research interests include the applications of game theory and queuing theory in wireless communication and networking and social networks. Dr. Jiang received the Best Paper Award from IEEE GLOBECOM in 2013, the Beijing Distinguished Graduated Student Award, Chinese National Fellowship and Tsinghua Outstanding Distinguished Doctoral Dissertation in 2013.
\end{IEEEbiography}

\begin{IEEEbiography}{Julian Cheng} (S'96, M'04, SM'13) received the B.Eng. degree (First Class) in electrical engineering from the University of Victoria, Victoria, BC, Canada in 1995, the M.Sc. (Eng.) degree in mathematics and engineering from Queen's University, Kingston, ON, Canada, in 1997, and the Ph.D. degree in electrical engineering from the University of Alberta, Edmonton, AB, Canada, in 2003. He is currently an Associate Professor in the School of Engineering, The University of British Columbia. His current research interests include digital communications over fading channels, orthogonal frequency division multiplexing, statistical signal processing for wireless applications, and optical wireless communications. Currently he serves as an Editor for IEEE COMMUNICATIONS LETTERS and IEEE TRANSACTIONS ON WIRELESS COMMUNICATIONS.
\end{IEEEbiography}

\begin{IEEEbiography}{Victor C. M. Leung} (S'75, M'89, SM'97, F'03) is a Professor of Electrical and Computer Engineering and holder of the TELUS Mobility Research Chair at the University of British Columbia (UBC).  His research is in the areas of wireless networks and mobile systems. He has co-authored more than 800 technical papers in archival journals and refereed conference proceedings, several of which had won best-paper awards. Dr. Leung is a Fellow of IEEE, a Fellow of the Royal Society of Canada, a Fellow of the Canadian Academy of Engineering and a Fellow of the Engineering Institute of Canada. He is serving or has served on the editorial boards of JCN, IEEE JSAC, Transactions on Computers, Wireless Communications, and Vehicular Technology, Wireless Communications Letters, and several other journals. He has provided leadership to the technical program committees and organizing committees of numerous international conferences. Dr. Leung was the recipient of the 1977 APEBC Gold Medal, NSERC Postgraduate Scholarships from 1977-1981, a 2012 UBC Killam Research Prize, and an IEEE Vancouver Section Centennial Award.
\end{IEEEbiography}

\begin{figure}[h]
        \centering
        \includegraphics*[width=12cm]{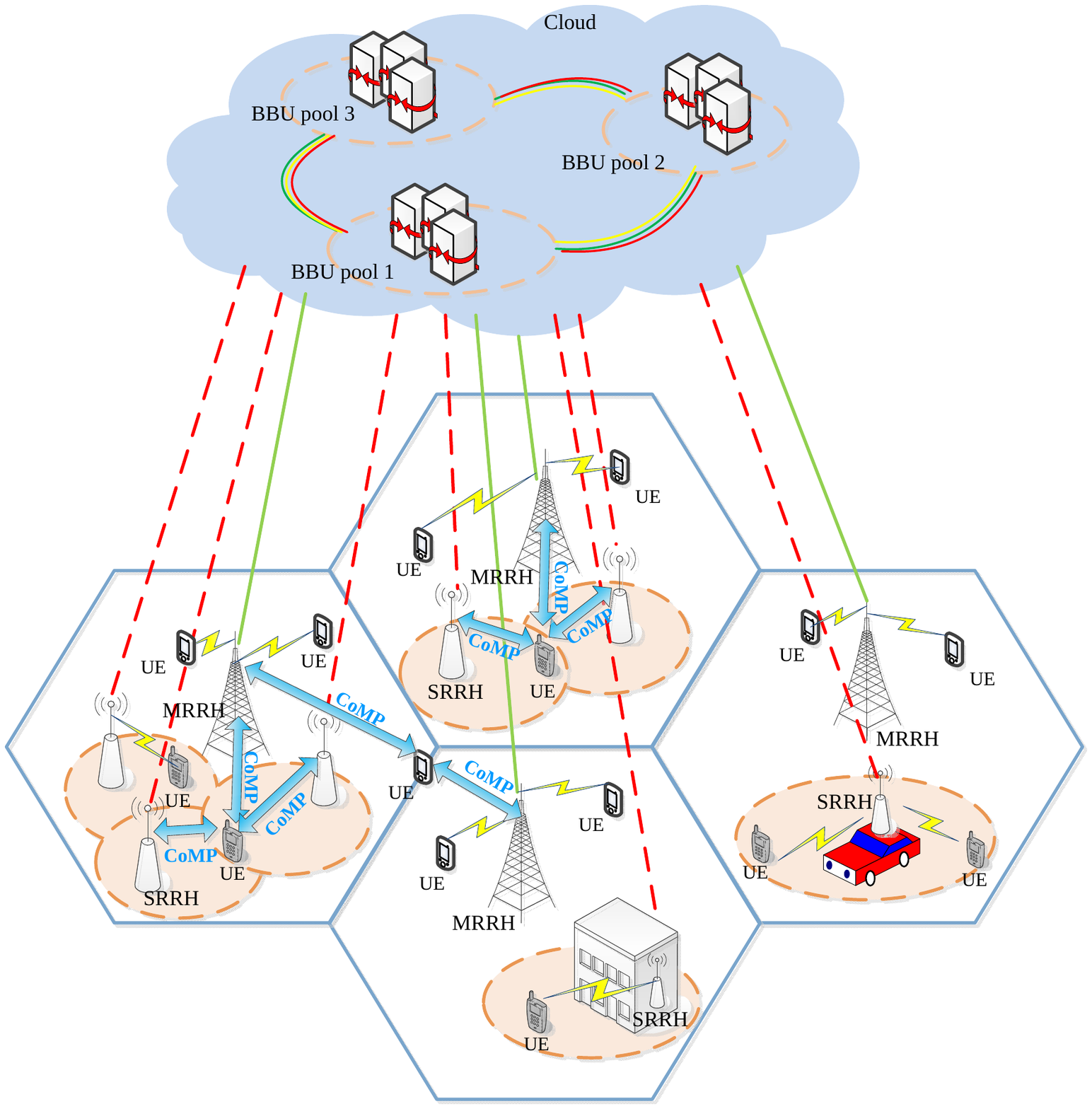}
        \caption{A HCSNet architecture.}
        \label{fig:1}
\end{figure}

\begin{figure}[h!]
        \centering
        \includegraphics*[width=16cm]{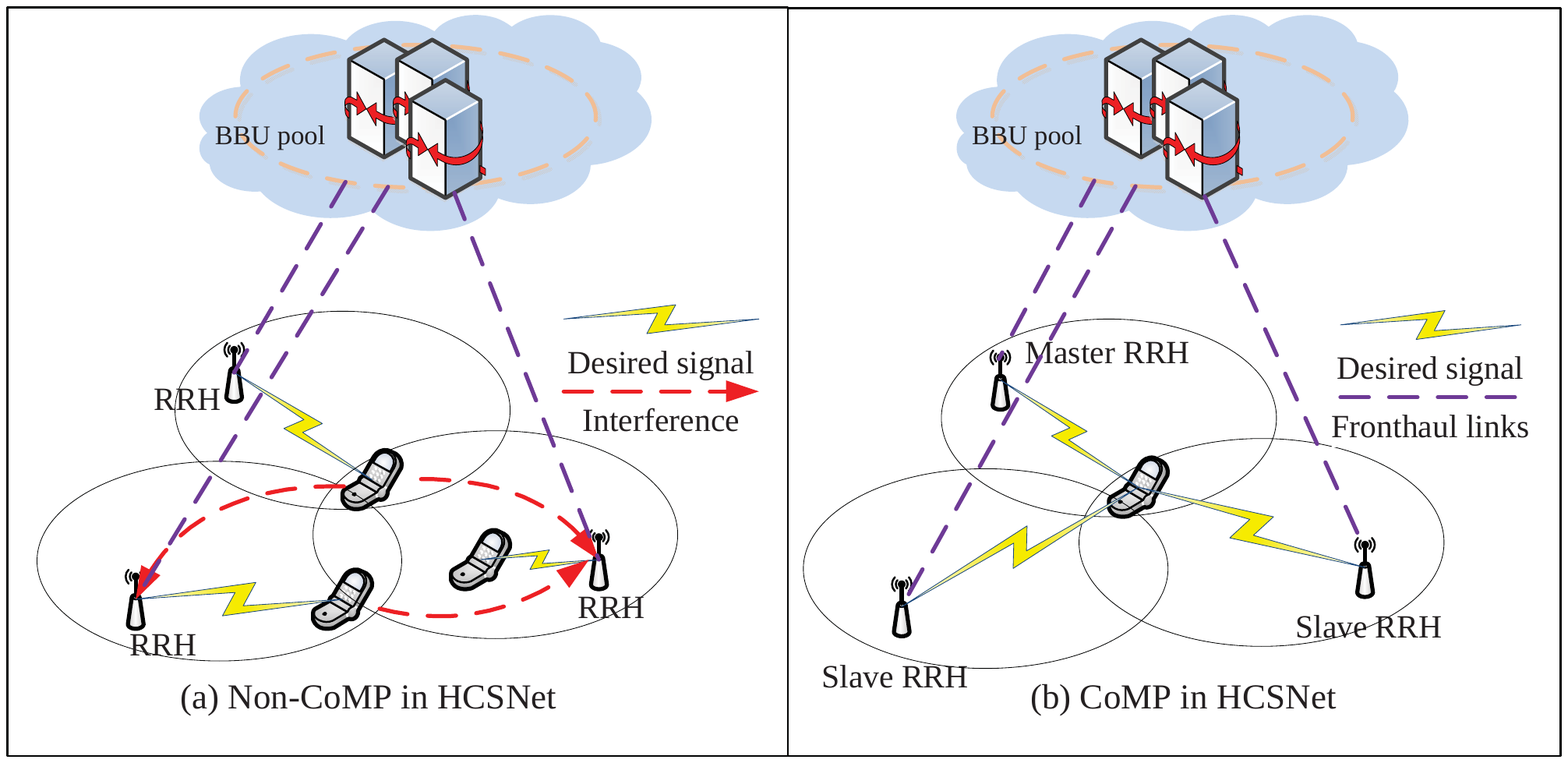}
        \caption{Non-CoMP and CoMP in HCSNet.}
        \label{fig:2}
\end{figure}

\begin{figure}[h!]
        \centering
        \includegraphics*[width=16cm]{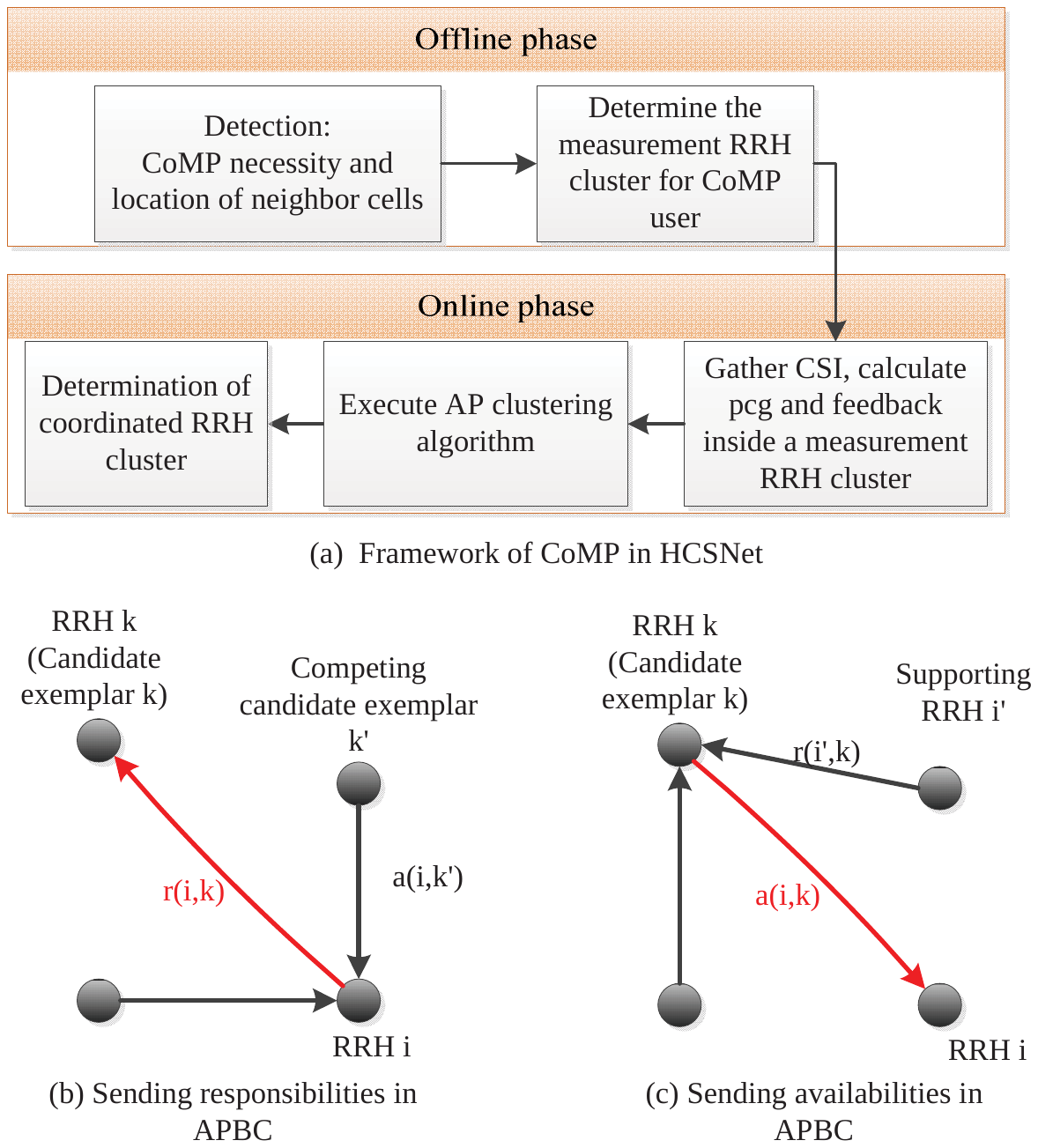}
        \caption{Framework of CoMP and diagram of responsibility and availability in HCSNet.}
        \label{fig:3}
\end{figure}

\begin{figure}[h]
        \centering
        \includegraphics*[width=16cm]{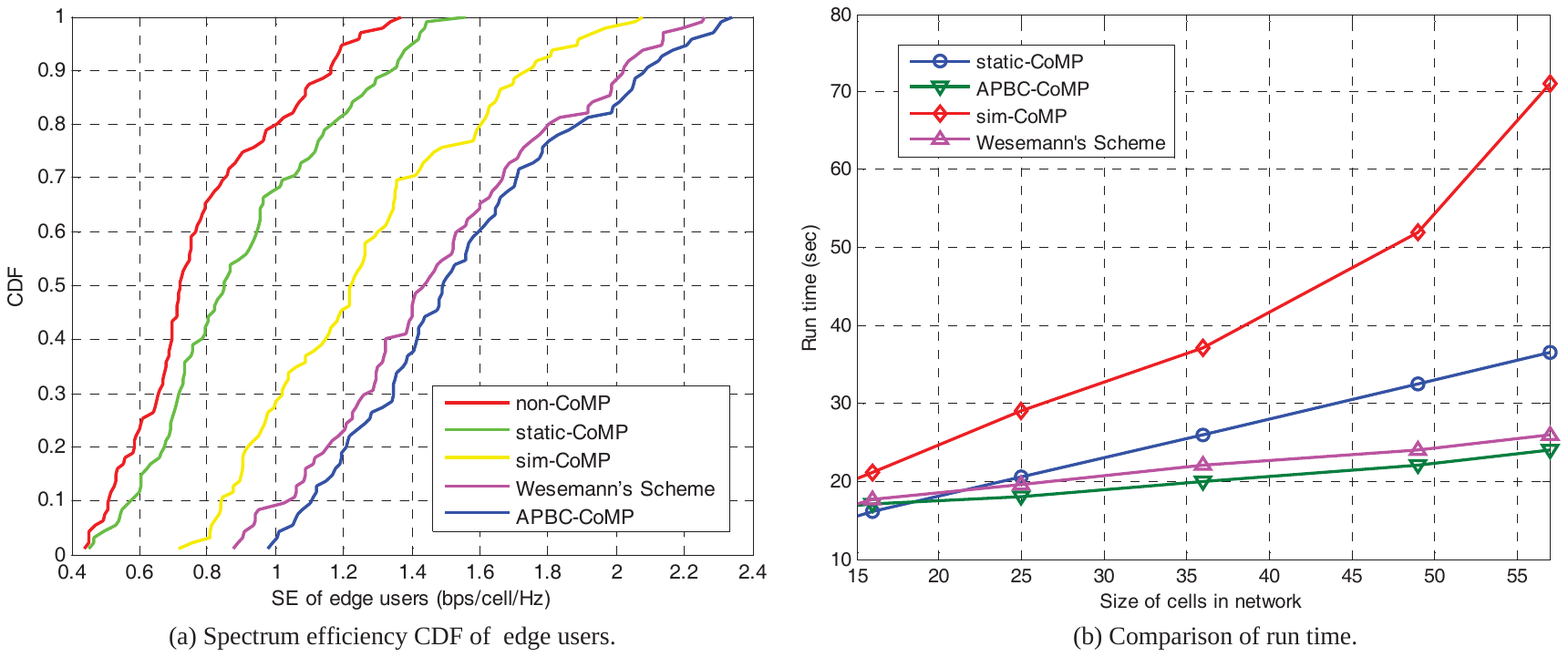}
        \caption{Evaluation of the presented CoMP scheme in HCSNet.}
        \label{fig:4}
\end{figure}

\begin{figure}[h]
        \centering
        \includegraphics*[width=15cm]{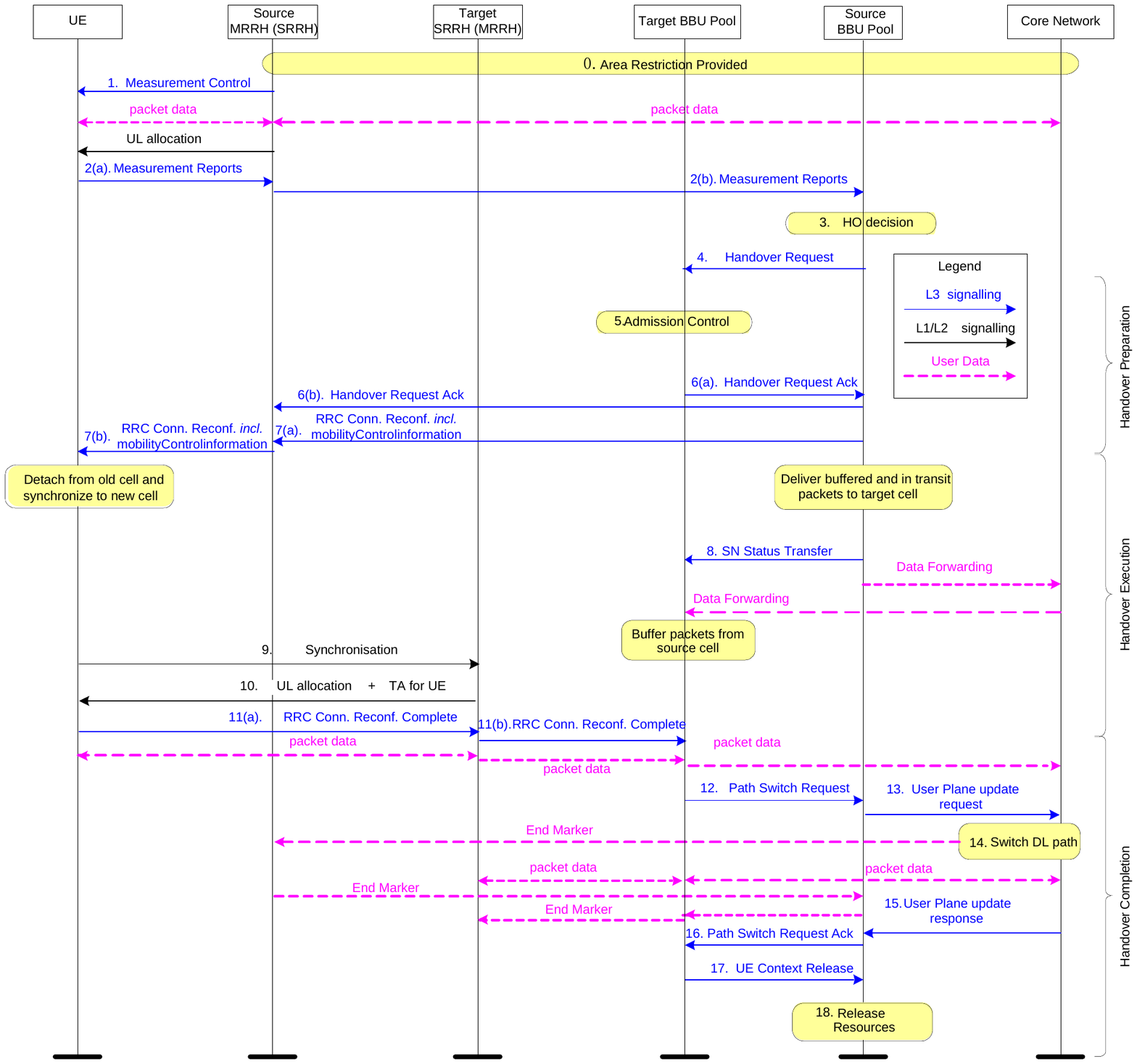}
        \caption{Handover procedure between SRRH and MRRH (Inter-BBU-pool) in HCSNet.}
        \label{fig:5}
\end{figure}

\begin{figure}[h]
        \centering
        \includegraphics*[width=16cm]{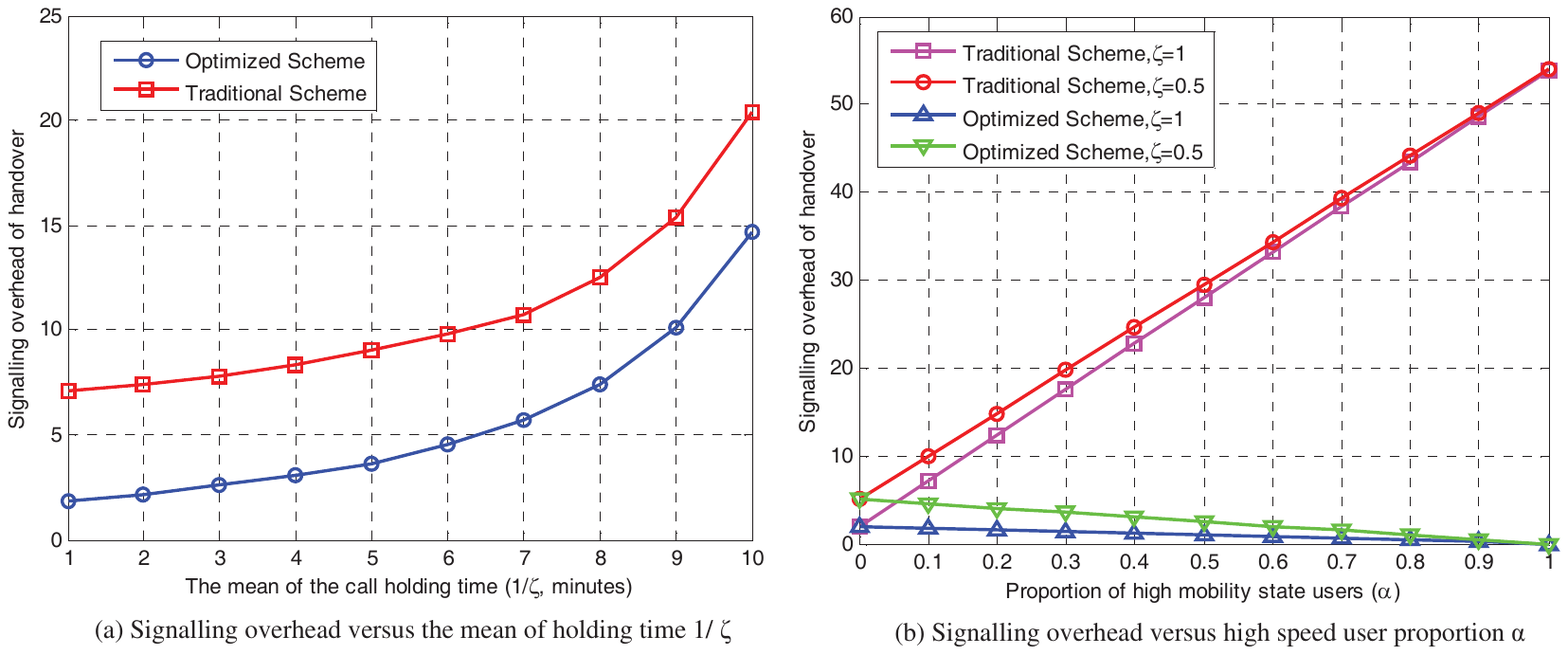}
        \caption{Signalling overhead of optimized scheme compared with existing scheme.}
        \label{fig:6}
\end{figure}

\end{document}